\documentclass[a4paper,aps,prd,10pt,preprintnumbers,showpacs,showkeys,twocolumn,superscriptaddress,nofootinbib,amsmath]{revtex4-1}
\usepackage{orcidlink}
\usepackage{graphicx,cmap}
\usepackage[utf8]{inputenc}
\usepackage[T1]{fontenc}
\usepackage{color}
\def\Order#1{{\cal O}\left(#1\right)}

\begin{document}

\title{Revisiting black holes in dark-matter halos: on consistent solutions to the Einstein equations}

\author{S. V. Bolokhov \orcidlink{0000-0002-9533-530X}
}
\email{bolokhov-sv@rudn.ru}
\affiliation{RUDN University, 6 Miklukho-Maklaya St, Moscow, 117198, Russian Federation}

\begin{abstract}
A number of recent papers have claimed to construct solutions of Einstein’s equations describing black holes surrounded by dark-matter halos with empirically motivated density profiles such as the Navarro--Frenk--White, Burkert, Einasto, pseudo-isothermal, and solitonic distributions. We show that {the approach used to obtain many of these metrics generically does not lead to the correct} solutions to the Einstein equations for the matter sources they purport to represent. {This issue} originates from applying the Newtonian relation between the tangential velocity and the enclosed mass directly within a relativistic framework, followed by the ad hoc assumption $g(r)=f(r)$ for the metric functions. This procedure leads to an anisotropic fluid with $P_r=-\rho$ and $P_t=-r\rho'/2-\rho$, whose density differs from the claimed halo profile and often becomes non-physical near the horizon, violating the weak energy condition.
As a result, the obtained spacetimes do not describe black holes embedded in known galactic halos but rather distinct anisotropic configurations unrelated to the intended matter distribution.  {We demonstrate this problem on} several representative examples from the literature, including metrics based on the NFW, Burkert, Einasto, solitonic, pseudo-isothermal, and Dehnen-(1,4,5/2) profiles, as well as the case of the NGC~4649 halo. For each case, the correct Einstein-consistent form of the metric and the associated physical interpretation are provided. Our analysis clarifies the limits of validity of the Newtonian approximation near compact objects and establishes a consistent framework for constructing dark-matter-inspired black-hole geometries within General Relativity.
\end{abstract}


\maketitle

\section{Introduction}

Astrophysical black holes do not exist in vacuum. On galactic and cluster scales they are embedded in extended dark-matter halos whose presence is inferred from stellar and gas rotation curves, lensing, and dynamical modeling. This basic fact has motivated a large number of recent works that attempt to construct self-consistent black-hole geometries surrounded by dark-matter distributions with phenomenologically motivated density profiles.

Consequently, extensive literature is devoted to self-consistent system of a supermassive black hole and galactic halo/dark matter and various effects in their background (see, for instance, \cite{Cardoso:2021wlq,Konoplya:2022hbl,Konoplya:2025nqv,Konoplya:2021ube,Konoplya:2025mvj,Jizba:2024owd,Konoplya:2025ect,Lutfuoglu:2025kqp,Fernandes:2025osu,Pathrikar:2025sin}). In parallel, there have been attempts to model specific observed systems, for example the giant elliptical galaxy NGC~4649 (M60), by combining a central black hole with the surrounding halo inferred from its kinematics~\cite{Shen:2009my,Lobo:2025kzb,Dubinsky:2025fwv}. In many such studies, the resulting metrics are taken as exact solutions of General Relativity, and are then used to discuss lensing, shadows, quasinormal ringing, accretion, and related observable effects.

A common strategy underlying these constructions can be traced to the approach proposed in Ref.~\cite{Matos:2003nb},
where the Newtonian relation between the tangential velocity and the enclosed mass was employed to approximate the relativistic lapse function. Subsequently, {for a generic metric  $ds^2=-f(r)dt^2+g(r)^{-1}dr^2+r^2 d\Omega^2$}, 
 Ref.~\cite{Xu:2018wow} introduced the additional assumption that the radial metric function equals the inverse of the lapse function, rather than deriving it from the Einstein equations using the appropriate mass function. This prescription has since been widely adopted in the literature.

The purpose of the present work is to show that, despite their widespread use, the resulting ``black hole in a halo'' metrics {are not exact solutions} to the Einstein equations with the matter content they are claimed to represent. The problem is twofold.

First, once the metric is fixed by requiring $f(r)=g(r)$ and tuned to reproduce a desired rotation curve, the resulting spacetime is no longer sourced by the original, advertised density profile $\widetilde{\rho}(r)$, but instead by an anisotropic fluid with stress--energy tensor
$$
P_r(r) = -\rho(r), 
\qquad 
P_t(r) = -\frac{r}{2}\rho'(r) - \rho(r),
$$
where $\rho(r)$ can be obtained {\it a posteriori} from the Einstein equations. 

Second, the same reconstruction machinery is regularly extrapolated down to radii close to the would-be event horizon, where the Newtonian circular-velocity argument used to define $m(r)$ is no longer valid. In that regime, the inferred radial pressure becomes large and negative,
%\begin{equation}
%P_r(r) = - \frac{m(r)^2}{2\pi r^4},
%\end{equation}
and can violate the weak energy condition near the horizon. This issue is particularly severe for small or moderate-mass black holes, for which the halo contribution is modelled as if it remained dilute and pressureless even deep in the strong-field region. Although it has been suggested that such an exotic stress tensor might in principle be mimicked by, e.g., a Bose--Einstein condensate and could even support wormholes~\cite{Jusufi:2019knb}, this interpretation relies on extending the Newtonian input well outside its domain of applicability. It is therefore not justified to interpret the resulting metric as a realistic black hole surrounded by an observed galactic halo.

In this work we revisit a number of representative examples in the literature and analyze them in a unified way. We show explicitly how the Matos--Núñez construction~\cite{Matos:2003nb} is commonly misapplied, and we identify the actual stress--energy tensor that sources the resulting metrics. We demonstrate that the widely used ``black hole in an NFW halo'' metric of Ref.~\cite{Xu:2018wow} is not sourced by an NFW density profile. The analogous statements apply to other density profiles in order to produce black-hole solutions, summarized in Table~\ref{tab:wrongbhhalo}.

Our goal is not only to point out these inconsistencies, but also to clarify what \emph{can} be done consistently. When the Einstein equations are solved with a specified density profile and with an explicitly anisotropic stress tensor, one indeed obtains static, spherically symmetric solutions that may describe a black hole surrounded by matter or even a fully regular geometry with a de~Sitter-like core. However, such solutions must then be interpreted for what they are: configurations supported by an anisotropic fluid with nontrivial radial and tangential pressures, rather than naive embeddings of a standard astrophysical halo around an otherwise classical black hole. This distinction is essential for any attempt to connect strong-field gravity with realistic galactic environments.

\begin{table*}
  \begin{tabular*}{\linewidth}{@{\extracolsep{\fill}}lll}
    Density distribution& $\widetilde{\rho}(r)$ & {Black-hole metrics for claimed density profile} \\
    \hline
    Navarro--Frenk--White \cite{Navarro:1996gj} & $\rho_0 r_s^3/r(r+r_s)^2$ & \cite{Xu:2018wow,Hou:2018bar,Zhang:2021bdr,Pantig:2022toh,Qin:2023nog,Capozziello:2023rfv,Ovgun:2023wmc,Tang:2023sig,Qiao:2024ehj,Kazempour:2024lcx,Zeng:2025kqw,Ahal:2025zxc,Li:2025eln} \\
    Burkert \cite{Burkert:1995yz,Salucci:2000ps} & $\rho_0 r_s^3/(r+r_s)(r^2+r_s^2)$ &  \cite{Jusufi:2019nrn,Zhang:2021bdr,Ovgun:2023wmc,Liu:2023xtb,Qiao:2024ehj} \\
    Moore \cite{Moore:1999gc} & $\rho_0 r_s^3/r^{3/2}(r+r_s)^{3/2}$ & \cite{Liu:2023xtb,Wu:2024hxr,Qiao:2024ehj,Li:2025eln} \\
    Einasto \cite{Retana-Montenegro:2012dbd} & $\rho_0 \exp\left(-\left(r/r_s\right)^{1/n}\right)$ & \cite{Liu:2023oab,Liu:2023vno} \\
    soliton solution \cite{Schive:2014dra} & $\rho_0 r_s^{16}/(r^2+r_s^2)^8$ & \cite{Pantig:2022sjb,DellaMonica:2023dcw} \\
    pseudo-isothermal \cite{Begeman:1991iy} & $\rho_0 r_s^2/(r^2+r_s^2)$ & \cite{Yang:2023tip} \\
    NGC~4649 (M60) \cite{Shen:2009my} & $\rho_0(r^2/3+r_s^2)/(r^2+r_s^2)^2$& \cite{Lobo:2025kzb} \\
    Dehnen-$(1, 4, \gamma)$ \cite{Dehnen:1993uh} & $\rho_0 r_s^4r^{-\gamma}(r+r_s)^{\gamma-4}$ & $\gamma=5/2$ \cite{Al-Badawi:2024asn,Liang:2025vux,Luo:2025xjb,Xamidov:2025prl}, $\gamma=0$ \cite{Pantig:2022whj,Toshmatov:2025rln}, $\gamma=1$ \cite{Pantig:2022whj,Jha:2025xjf,Toshmatov:2025rln}, $\gamma=2$ \cite{Uktamov:2025lwb,Pathrikar:2025sin} \\
    $\beta$-model \cite{Navarro:1994hi} & $\rho_0 r_s^3/(r^2+r_s^2)^{3/2}$ & \cite{Liu:2023xtb,Wu:2024hxr,Qiao:2024ehj,Li:2025eln} \\
    dark-matter spike \cite{Tang:2020jhx} & $\rho_0(1-M/r)^3(r_s/r)^{\alpha}$ & \cite{Xu:2021dkv,Zhang:2021bdr,Capozziello:2023rfv,Liu:2024xcd}\\
    Brownstein \cite{Brownstein:2009gz} & $\rho_0 r_s^3/(r^3+r_s^3)$ & \cite{Liu:2023xtb} \\
    \hline
  \end{tabular*}
  \caption{Summary of the papers, which present black holes surrounded by the halo. {The 3rd column lists the papers considering the black-hole metrics within the approach discussed, which generally does not lead to exact solutions to Einstein's equations for a claimed density distribution.} Constants $\rho_0$ and $r_s$ are the characteristic density and scale parameter.}
  \label{tab:wrongbhhalo}
\end{table*}

\section{Matos-Núñez metric due to the spherically symmetric halo}\label{sec:MatosNunez}

We start from the general line element for a spherically symmetric spacetime,
\begin{eqnarray}\label{line-element}
ds^2&=&-f(r)dt^2+\frac{dr^2}{g(r)}+r^2d\Omega^2,
\\\nonumber
&&d\Omega^2=d\theta^2+\sin^2\theta d\varphi^2.
\end{eqnarray}

Conventionally, the right-hand side of the Einstein equations is attributed to a matter distribution characterized by the energy density $\rho(r)$, radial pressure $P_r(r)$, and tangential pressure $P_t(r)$:\footnote{{Here we use the system of units $G=c=1$.}}
\begin{eqnarray}
\label{Gtt}
&&\frac{g(r)-1+rg'(r)}{r^2}=8\pi T^t_t=-8\pi \rho(r),\\
\label{Grr}
&&\frac{g(r)-1}{r^2}+\frac{g(r)}{f(r)}\frac{f'(r)}{r}=8\pi T^r_r=8\pi P_r(r),\\
&&\frac{g'(r)}{2}\left(\frac{1}{r}+\frac{f'(r)}{2f(r)}\right)+\frac{g(r)}{2f(r)}\left(f''(r)+\frac{f'(r)}{r}-\frac{f'(r)^2}{2f(r)}\right)
\nonumber\\
&&\hspace{8.8em} =8\pi T^{\theta}_{\theta}=8\pi T^{\varphi}_{\varphi}=8\pi P_t(r).\label{Gaa}
\end{eqnarray}

It is convenient to introduce the mass function $m(r)$ as
\begin{equation}\label{gmass}
    g(r)=1-\frac{2m(r)}{r},
\end{equation}
so that Eq.~(\ref{Gtt}) reads
\begin{equation}\label{massfunction}
    m'(r)=4\pi r^2 \rho(r).
\end{equation}

Within General Relativity, determining the metric functions requires specifying the right-hand side of the Einstein equations. However, for a dark-matter halo, the equation of state is unknown. Therefore, Ref.~\cite{Matos:2003nb} adopted a different approach based on observational data for the tangential velocity of stars in galaxies. Specifically, the mass function was inferred from the centrifugal force in the Newtonian approximation at radius $r$,
\begin{equation}\label{tangential}
    \frac{m(r)}{r}=v_{tg}^2.
\end{equation}
Once $m(r)$ is known, $g(r)$ follows from Eq.~(\ref{gmass}), while the lapse function $f(r)$ can be obtained using the relativistic expression for the tangential velocity at the same circular orbit,
\begin{equation}\label{tangentiallapse}
    v_{tg}^2=\frac{r}{2}\frac{f'(r)}{f(r)}. %=\frac{d\ln\sqrt{f(r)}}{d\ln r}.
\end{equation}
Hence,
\begin{equation}\label{masslapse}
    \frac{f'(r)}{f(r)}=\frac{2m(r)}{r^2}.
\end{equation}

Once the mass function is obtained from Eq.~(\ref{tangential}), the corresponding density profile can be found using Eq.~(\ref{massfunction}).

Substituting {Eqs.~(\ref{gmass}) and (\ref{masslapse})} into Eq.~(\ref{Grr}) yields the radial pressure
\begin{equation}\label{eqofstate}
2\pi P_r(r)=-\frac{m(r)^2}{r^4}.
\end{equation}
Similarly, from Eq.~(\ref{Gaa}) one can derive the tangential pressure $P_t(r)$. These expressions do not stem from any fundamental theory but are instead a consequence of adopting the Newtonian relation (\ref{tangential}) within the relativistic framework.

As discussed in detail in Ref.~\cite{Matos:2003nb}, the resulting stress-energy tensor is valid only in the regime of very low densities, where the corresponding pressures are negligibly small and the Newtonian approximation remains applicable. In contrast, for compact configurations, such as small black holes, the negative radial pressure can become arbitrarily large near the event horizon and violate the weak energy condition. 

Nevertheless, Ref.~\cite{Xu:2018wow} applied this method to obtain a metric purported to describe a black hole embedded in a dark-matter halo with the Navarro--Frenk--White (NFW) density profile~\cite{Navarro:1996gj},
\begin{equation}\label{NFWdensity}
    \widetilde{\rho}(r) = \frac{\rho_c r_s^3}{r(r+r_s)^2}.
\end{equation}
Using the mass function obtained by integrating {Eq.~(\ref{massfunction})}, the lapse function was determined via Eq.~(\ref{masslapse}), followed by the additional assumption $g(r) = f(r)$. It is straightforward to verify that this condition corresponds to an anisotropic fluid with
\begin{equation}\label{vacuumequation}
  P_r(r)=-\rho(r), \qquad P_t(r)=-\frac{r}{2}\rho'(r)-\rho(r),
\end{equation}
where $\rho(r)$ differs from the NFW density $\widetilde{\rho}(r)$ in Eq.~(\ref{NFWdensity}). The reason is that the metric functions in {equation (4)} of Ref.~\cite{Xu:2018wow},
\begin{equation}\label{Xu}
    f(r)=g(r)=\left(1+{\frac{r}{r_s}}\right)^{-\alpha/r}, \quad {\alpha=8\pi \rho_c r_s^3,}
\end{equation}
were derived without direct reference to the Einstein equations. Substituting Eq.~(\ref{Xu}) into Eq.~(\ref{Gtt}) gives
%\begin{eqnarray}\label{effectiverho}
%\rho(r)&=&\frac{1}{8 \pi  r^2}\Biggl(1-\left(1+\frac{r_s}{r}\right)^{-\alpha /r}\\\nonumber&-&\alpha \left(1+\frac{r_s}{r}\right)^{{-1-\alpha/r}}\frac{(r+r_s) \ln\left(1+r_s/r\right)+r_s}{r^2}\Biggr),
%\end{eqnarray}
\begin{multline}\label{effectiverho}
\rho(r)=\frac{1}{8\pi r^2}\Biggl(
1-\Bigl(1+\frac{r}{r_s}\Bigr)^{\!\!-\frac{\alpha}{r}} \\
-\alpha\Bigl(1+\frac{r}{r_s}\Bigr)^{\!\!-\frac{\alpha}{r}}
\;\frac{\left(r+r_s\right) \ln\left(1+{r}/{r_s}\right)-r}{r\left(r+r_s\right)}
   \Biggr),
\end{multline}
which in the limit of low density approaches a different profile
$$\rho(r)=\frac{\alpha}{8\pi r^2(r+r_s)}+\Order{\alpha}^2.$$

Finally, the metric with a horizon obtained in Ref.~\cite{Xu:2018wow} arises after several manipulations that effectively reduce to the following procedure:
take the density (\ref{effectiverho}), solve Eq.~(\ref{massfunction}) for the mass function $m(r)$, and interpret the integration constant $M$ as the black-hole mass. The resulting lapse function remains nonsingular only under the so-called cold dark matter conditions (\ref{vacuumequation}). In this case, the metric functions take the form
\begin{equation}\label{XuBH}
    f(r)=g(r)=\left(1+\frac{r}{r_s}\right)^{-\alpha/r}-\frac{2M}{r}.
\end{equation}

Thus, the metric functions (\ref{XuBH}) satisfy the Einstein equations with the anisotropic fluid defined by Eq.~(\ref{vacuumequation}) and the density profile~(\ref{effectiverho}), which differs from the claimed NFW profile (\ref{NFWdensity}). {Therefore, a combination of the Newtonian approximation and the general relativistic framework, represented in Ref.~\cite{Xu:2018wow}, cannot be regarded as a consistent approach.} Further, the same {controversial} technique has been used in various papers to produce the black-hole solutions embedded in the dark-matter halo (see Table~\ref{tab:wrongbhhalo} for summary).

\section{NGC 4649 halo}

{In the interesting paper~\cite{Lobo:2025kzb}, two black-hole metrics (the so-called ``first solution'' and ``second solution'') were proposed} for the dark-matter halo of NGC~4649 (M60)~\cite{Shen:2009my}, described by the density profile
\begin{equation}\label{NGCdensity}
 \rho(r)=\frac{V_c^2}{4\pi}\frac{3a^2+r^2}{(r^2+a^2)^2}.
\end{equation}
The first solution was obtained via the approach of~\cite{Xu:2018wow} considered above, and suffers from the same issue: it does not reproduce an exact solution to the Einstein equations with the density distribution (\ref{NGCdensity}).

For the second solution the mass function is derived by integrating Eq.~(\ref{NGCdensity}), and the Einstein equations are solved under the conditions (\ref{vacuumequation}). This yields the metric functions\footnote{Note that the solution in Ref.~\cite{Lobo:2025kzb} contains another sign.}
\begin{equation}\label{solution-metric}
  f(r)=g(r)=1-\frac{2M}{r}-\frac{2V_c^2r^2}{a^2+r^2}
\end{equation}
which reproduce the density profile \eqref{NGCdensity}. Here $M$ is an arbitrary integration constant.
For $M = 0$, the metric exhibits a de~Sitter core:
$$f(r)=g(r)=1-\frac{2V_c^2}{a^2}r^2+\Order{r}^4,$$
and is therefore regular at $r = 0$. For sufficiently large halo scale velocity, $V_c^2 \geq 1/2$, the metric possesses a cosmological horizon at
\begin{equation}
r_c=\frac{a}{\sqrt{2V_c^2-1}}.
\end{equation}

However, realistic estimates of the halo scale velocity~\cite{Shen:2009my} are several orders of magnitude smaller than the speed of light, typically $V_c \sim 10^{-3} \ll 1$. In this regime, the metric function (\ref{solution-metric}) describes a black hole of mass $M$ embedded in a dark-matter halo. The total mass of the configuration remains finite only if the density profile~(\ref{NGCdensity}) is valid within a finite spatial region, $r < r_m$. The total mass of the black hole and halo is then
$$M_{tot}=\frac{V_c^2r_m^3}{a^2+r_m^2}+M.$$

It is interesting to note that the metric (\ref{solution-metric}) remains asymptotically flat in the limit of $r_m\to\infty$. The resulting configuration behaves asymptotically as the global monopole spacetime with a deficit solid angle \cite{Barriola:1989hx}. The time of the asymptotic observer $T$ is related to the coordinate
\begin{equation}
  dT=dt\sqrt{1-2V_c^2}.
\end{equation}
It is possible to introduce the generalized ADM mass \cite{Nucamendi:1996ac}
\begin{equation}
  \widetilde{M}=\frac{M}{(1-2V_c^2)^{3/2}},
\end{equation}
and the line element (\ref{line-element}) takes the following form:
\begin{eqnarray}\label{line-element-deficit}
  ds^2&=&-\widetilde{f}(R)dT^2+\frac{dR^2}{\widetilde{f}(R)}+(1-2V_c^2)R^2d\Omega^2,
  \\\nonumber
  &&R\equiv\frac{r}{\sqrt{1-2V_c^2}},
\end{eqnarray}
where
\begin{equation}\label{asymptotictimemetric}
\widetilde{f}(R)=1-\frac{2\widetilde{M}}{R}-\frac{2V_c^2a^2/(1-2V_c^2)}{a^2+R^2-2V_c^2R^2}.
\end{equation}

\section{Dehnen-(1,4,5/2) profile}

Another example of a controversial technique considered in Sec. \ref{sec:MatosNunez} can be found in Ref.~\cite{Al-Badawi:2024asn}. The paper begins with the Dehnen-$(1,4,5/2)$ density profile~\cite{Dehnen:1993uh},
\begin{equation}\label{Dehnen1452density}
    \widetilde{\rho}(r)=\frac{\rho_s r_s^4}{r^{5/2}(r+r_s)^{3/2}}.
\end{equation}
Integrating Eq.~(\ref{massfunction}) yields the mass function
\begin{equation}\label{Dehnen1452massfunc}
    m(r)=8\pi\rho_sr_s^3\sqrt{\frac{r}{r+r_s}},
\end{equation}
which is then used to obtain the lapse function via {solving} Eq.~(\ref{masslapse}),
\begin{equation}
    f(r)=\exp\left(A-32\pi\rho_sr_s^2\sqrt{1+\frac{r_s}{r}}\right),
\end{equation}
where $A$ is an integration constant that must be fixed to ensure an asymptotically Minkowski geometry, yielding $A = 32\pi\rho_s r_s^2$. {Note that the authors' choice $A = 0$ leads} to a solution with a solid-angle deficit,
\begin{equation}
    f(r)=\exp\left(-32\pi\rho_sr_s^2\sqrt{1+\frac{r_s}{r}}\right).
\end{equation}

Next, the authors invoke the low-density approximation for the halo matter,
\begin{equation}
    f(r)=1-32\pi\rho_sr_s^2\sqrt{1+\frac{r_s}{r}}+\Order{\rho_s}^2.
\end{equation}

Following the same manipulations described in Sec.~\ref{sec:MatosNunez}, they obtain the following black-hole metric:
\begin{equation}\label{wrongBHmetric}
    f(r)=g(r)=1-32\pi\rho_sr_s^2\sqrt{1+\frac{r_s}{r}}-\frac{2M}{r}.
\end{equation}

Substituting Eq.~(\ref{wrongBHmetric}) into Eq.~(\ref{Gtt}) yields the corresponding density distribution,
\begin{equation}
    \rho(r)=\frac{2\rho_s r_s^2(2r+r_s)}{r^{5/2}(r+r_s)^{1/2}},
\end{equation}
which is neither of the Dehnen type nor reproduces the profile~(\ref{Dehnen1452density}) in any limit.

The correct solution of the Einstein equations with the density profile~(\ref{Dehnen1452density}) and the additional condition~(\ref{vacuumequation}) is
\begin{equation}\label{correctdehnenmetric}
    f(r)=g(r)=1-\frac{2M}{r}-\frac{16\pi\rho_sr_s^3}{\sqrt{r(r+r_s)}},
\end{equation}
which asymptotically approaches flat spacetime without any angular deficit.

\section{A simple probe of the black hole geometry: Eikonal quasinormal modes}\label{sec:eikonal}

In the high-multipole limit, $\ell \gg 1$, the quasinormal spectrum of a broad class of spherically symmetric black holes admits a remarkably simple geometric interpretation. In this regime, the dominant contribution to the effective potential governing linear perturbations comes from the centrifugal barrier,
\begin{equation}
    V(r) \simeq f(r)\,\frac{\ell(\ell+1)}{r^{2}},
\end{equation}
where $f(r)=g(r)$ is the metric function in the line element.
Since the potential is sharply peaked at large $\ell$, the eikonal quasinormal frequencies are determined by the properties of the unstable circular null geodesics of the background spacetime.

Radial motion of a massless particle with conserved energy $E$ and angular momentum $L$ can be written as
\begin{equation}
    \dot r^{2} + V_{\rm geo}(r) = E^{2}, \qquad 
    V_{\rm geo}(r) = f(r)\frac{L^{2}}{r^{2}} ,
\end{equation}
where dots denote derivatives with respect to an affine parameter.  
A circular null orbit at radius $r=r_{c}$ satisfies
\begin{equation}\label{eq:geocond}
    V_{\rm geo}(r_{c}) = E^{2}, \qquad 
    \frac{d V_{\rm geo}}{dr}\bigg|_{r_{c}} = 0 .
\end{equation}
The first condition determines the ratio $E/L$, while the second yields the photon-sphere equation
\begin{equation}\label{eq:phs}
    2f(r_{c}) - r_{c} f'(r_{c}) = 0 ,
\end{equation}
whose positive root gives the radius of the unstable circular null geodesic.

The orbital (Keplerian) frequency measured at infinity is
\begin{equation}\label{eq:Omega_c}
    \Omega_{c} = \frac{\sqrt{f(r_{c})}}{r_{c}} .
\end{equation}
Small radial perturbations around $r=r_{c}$ grow exponentially with a rate given by the Lyapunov exponent~\cite{Cardoso:2008bp}
\begin{equation}\label{eq:lambda}
    \lambda = \sqrt{-\frac{V_{\rm geo}''(r_{c})}{2\dot t^{2}}}
    = \sqrt{\frac{f(r_{c})}{2r_{c}^{2}}
            \bigg[ f(r_{c}) - r_{c}^{2} f''(r_{c})/2 \bigg] } .
\end{equation}
The general result of \cite{Cardoso:2008bp} establishes that, for any static and spherically symmetric black hole whose perturbation equations possess a standard single-barrier form, the quasinormal spectrum at $\ell \gg 1$ is
\begin{equation}\label{eq:eikonalQNM}
    \omega_{\ell n} 
      = \left(\ell+\frac{1}{2}\right)\, \Omega_{c}
        - i\left(n+\frac12\right)\lambda 
        + \Order{\frac{1}{\ell}},
\end{equation}
where $n=0,1,2,\dots$ denotes the overtone number.  
Thus, the real part of the quasinormal frequency is governed by the angular velocity of the circular null orbit, whereas the imaginary part is controlled by the instability timescale of this orbit encoded in $\lambda$.

The above result can also be derived using the WKB method, originally proposed for computing black-hole quasinormal modes in \cite{Schutz:1985km} and later extended to higher orders in \cite{Iyer:1986np,Konoplya:2003ii,Matyjasek:2017psv}. The WKB approximation is generally accurate for the least damped modes at small and moderate $\ell$, and becomes highly precise in the eikonal limit $\ell \to \infty$ (see ~\cite{Malik:2024qsz,Dubinsky:2024aeu,Cuyubamba:2016cug,Zhao:2023tyo,Konoplya:2024lch,Dubinsky:2024gwo,Malik:2024nhy,Abdalla:2005hu,Aneesh:2018hlp,Kodama:2009bf,Skvortsova:2023zca,Konoplya:2013sba,Zhao:2023itk,Malik:2024elk,Stuchlik:2025mjj,Konoplya:2020hyk,Dubinsky:2024mwd,Churilova:2021tgn,Qian:2022kaq,Zhidenko:2003wq,Konoplya:2025uiq,Arbelaez:2025gwj,Pathrikar:2025gzu,Skvortsova:2024wly,Malik:2025erb,Bolokhov:2025lnt} for examples). Nevertheless, this statement must be treated with caution, since a number of counterexamples are known in which the correspondence fails \cite{Konoplya:2017wot,Konoplya:2022gjp,Bolokhov:2023dxq,Konoplya:2025afm,Konoplya:2020bxa,Khanna:2016yow}. In the present case, however, the effective potential exhibits standard WKB-friendly behavior, so the eikonal regime is reliable, and the WKB method correctly reproduces the null-geodesic characteristics derived above.

Using the metric function (\ref{wrongBHmetric}), one can find
\begin{eqnarray}\nonumber
\Omega_c&=&\frac{1}{3\sqrt{3}M}\left(1-48\pi\rho_s r_s^2\sqrt{1+\frac{r_s}{3M}}\right)+\Order{\rho_s}^2,
\\\nonumber
\lambda&=&\frac{1}{3\sqrt{3}M}\left(1-\frac{64\pi\rho_s r_s^2\left(1+\dfrac{5r_s}{8M}+\dfrac{29r_s^2}{288M^2}\right)}{\left(1+\dfrac{r_s}{3M}\right)^{3/2}}\right)
\\\nonumber&&+\Order{\rho_s}^2,
\end{eqnarray}
which neither reproduces the Schwarzschild limit for $M\gg r_s$,
whereas for the correct solution (\ref{correctdehnenmetric}), the eikonal expression takes the form
\begin{eqnarray}\nonumber
\Omega_c&=&\frac{1}{3\sqrt{3}M}\left(1-\frac{8\pi\rho_s r_s^3}{M\sqrt{1+\dfrac{r_s}{3M}}}\right)+\Order{\rho_s}^2,
\\\nonumber
\lambda&=&\frac{1}{3\sqrt{3}M}\left(1-\frac{8\pi\rho_s r_s^3\left(1+\dfrac{7r_s}{9M}+\dfrac{29r_s^2}{216M^2}\right)}{M\left(1+\dfrac{r_s}{3M}\right)^{5/2}}\right)
\\\nonumber&&+\Order{\rho_s}^2.
\end{eqnarray}
Thus, the non-self-consistent approach leads to the substantially different observables, and, therefore, cannot be used even as an approximate representation of a genuine solution of the Einstein equations. The above eikonal expressions for quasinormal modes can be extended further beyond the eikonal regime using the general approach developed in \cite{Konoplya:2023moy}.

\section{Conclusions}

In this work we have revisited a number of metrics proposed in the recent literature as models of black holes surrounded by dark-matter halos and have shown that many of them do not satisfy the Einstein equations with the density profiles they claim to represent. The common problem originates from applying the Newtonian relation between the tangential velocity and enclosed mass within the relativistic framework and from the additional assumption $f(r)=g(r)$ imposed without justification. As a result, the stress--energy tensors sourcing those spacetimes correspond to anisotropic fluids with negative radial pressure and densities that differ substantially from the originally assumed halo profiles, sometimes even becoming nonphysical near the horizon.

For some of the density profiles we present the correct black-hole solutions for the first time.
The availability of such exact metrics opens the way for consistent studies of a wide range of physical phenomena.  
In particular, one can now analyze quasinormal modes and grey-body factors of test fields propagating in these spacetimes, compute the corresponding ringdown spectra, and investigate possible observational imprints of the surrounding halo on black-hole oscillations.  
The same geometries can be used to calculate gravitational lensing, shadow profiles, and photon-sphere properties in a fully self-consistent manner, providing reliable theoretical templates for comparison with current and future observations.  
Furthermore, because the corrected configurations interpolate smoothly between vacuum black holes and matter-supported solutions, they may also serve as a basis for exploring regular black holes and wormhole-like transitions in dark-matter environments.

In summary, the present analysis not only clarifies some inconsistencies in a large body of recent literature but also provides a consistent and physically motivated framework for constructing and studying black-hole spacetimes immersed in realistic halo models.  
The metrics Eqs.~(\ref{asymptotictimemetric}) and~(\ref{correctdehnenmetric}) presented in this work can be regarded as the proper starting point for any future investigation of dynamical, optical, or thermodynamical phenomena in dark-matter-inspired black-hole backgrounds, including, first of all, quasinormal modes \cite{Konoplya:2011qq,Kokkotas:1999bd,Bolokhov:2025uxz}, grey-body factors \cite{Konoplya:2019hlu} and shadows \cite{Perlick:2021aok,Konoplya:2019sns}.

\bibliography{manuscript}

\end{document}